\documentclass[useAMS,usenatbib]{mn2e}
\usepackage{graphicx,amssymb, txfonts}

\def\br{Br$\gamma$}

\def\h2{H$_2$}
\def\p1{Paper~I}

\def\str{{\sc starlight}}
\def\mc{$\mu$m} 

\title[Intermediate age stars and low-$\sigma_*$ ring in Mrk\,1157]{Intermediate-age Stars as Origin of Low Stellar velocity Dispersion Nuclear Rings: the case of Mrk\,1157}
\author[Riffel et al.]{Rog\'erio Riffel$^{1}$\thanks{E-mail:
riffel@ufrgs.br}, Rogemar A. Riffel$^{2}$,  Fabricio Ferrari$^3$, Thaisa Storchi-Bergmann$^{1}$\\
1. Universidade Federal do Rio Grande do Sul, Instituto de F\'\i sica, CP 15051, 91501-970, Porto Alegre, RS, Brazil.\\
2. Universidade Federal de Santa Maria, Departamento de F\'\i sica, Centro de Ci\^encias Naturais e Exatas, 
97105-900, Santa Maria, RS, Brazil.\\
3. Universidade Federal de Rio Grande, Instituto de Matem\'atica Estat\'{i}stica e F\'{i}sica, CP 474, 96201-900, Rio Grande, RS, Brazil.\\
}
\begin{document}

\date{}

\pagerange{\pageref{firstpage}--\pageref{lastpage}} \pubyear{2002}

\maketitle

\label{firstpage}

\begin{abstract}
  We have used the Gemini Near-infrared Integral Field Spectrograph (NIFS) to
  map the age distribution of the stellar population in the inner 400\,pc of the
  Seyfert 2 galaxy Mrk\,1157 (NGC\,591), at a spatial resolution of 35\,pc. We
  have performed wavelet and principal component analysis in the data in order
  to remove instrumental signatures.  An old stellar population component
  (age $\gtrsim$5\,Gyr) is dominant within the inner $\approx$\,130\,pc which we
  attribute to the galaxy bulge. Beyond this region, up to the borders of the
  observation field, young to intermediate age components (0.1--0.7\,Gyr)
  dominate.  As for Mrk\,1066, previously studied by us, we find a spatial
  correlation between this intermediate age component and a partial ring of low
  stellar velocity dispersions ($\sigma_*$).  Low-$\sigma_*$ nuclear rings have
  been observed in other active galaxies and our results for Mrk\,1157 and
  Mrk\,1066 reveal that they are formed by intermediate age stars. Such age is
  consistent with a scenario in which the origin of the low-$\sigma_*$ rings is
  a past event which triggered an inflow of gas and formed stars which still
  keep the colder kinematics of the gas from which they have formed.  No
  evidence for the presence of an unresolved featureless continuum and hot dust
  component -- as found in Mrk1066 -- are found for Mrk\,1157.
\end{abstract}


\section{Introduction}


The co-existence of an active galactic nucleus (AGN) and young stars in the
central region of Seyfert galaxies is a widely known phenomenon,
\citep[e.g.][]{sb00,sb01,gd01,cid04,asari07,dors08}, supporting the so-called
AGN-Starburst connection
\citep[e.g.][]{norman88,terlevich90,heckman97,heckman04,rogemar09c}.  The above
studies have pointed out that the main difference between the stellar population
(SP) of active and non-active galaxies is an excess of intermediate age stars in
the former.  In addition, near-infrared (NIR) SP studies have revealed that the
continuum is also dominated by the contribution of intermediate-age stellar
population components \citep[SPCs,][]{rogerio07,rogerio09,rogemar10b,martins10}.
In the NIR, another component is commonly detected in the nuclear spectra of Seyfert galaxies:
unresolved hot dust emission \citep{rogemar09b,rogemar09c,rogerio09,ardila06,ardila05}. Thus, the study of the contribution
of the SPs and other components to the circumnuclear continuum of active
galaxies is a fundamental key in the understanding of the nature of their
central engine.

Very recently, the use of integral field spectroscopy with adaptive optics at
the Gemini North Telescope has allowed us to derive the contribution of distinct
SPCs to the NIR spectra of active galaxies. In addition, we have mapped SPCs spatial distributions and
performed the first two-dimensional (2D) SP synthesis in the NIR of the nuclear
region of an active galaxy \citep[Mrk\,1066, ][]{rogemar10b}. A spatial
correlation between the intermediate age SPC and a partial ring of low stellar
velocity dispersions ($\sigma_*$) was found, supporting the interpretation that
the low-$\sigma_*$ structures commonly observed in the inner few hundreds 
parsecs of Seyfert galaxies are due to colder regions with more recent star
formation than the underlying bulge
\citep{barbosa06,rogemar08,rogemar09a,rogemar10d}. Using a different method ---
modelling the \br\ equivalent width, supernova rate and mass-to-light ratio ---
\citet{davies07} have quantified the star formation history in the centre of 9
nearby Seyfert galaxies using their {\sc stars} code.  They found that the ages
of the stars which contribute most to the NIR continuum lie in the range
10--30\,Myr, pointing out that these ages should be considered only as
``characteristic", as they have not performed a proper spectral synthesis,
suggesting that there may be simultaneously two or more SPs that are not coeval
\citep{davies07,davies06}.

As part of an ongoing project aimed at mapping the 2D age distribution of the
NIR SP in the inner few hundred parsecs of Seyfert galaxies, we present the
NIR SP synthesis of the Seyfert 2 galaxy Mrk\,1157 (NGC\,591), an early type
barred spiral galaxy located at a distance $d=61.1$\,Mpc (from NASA/IPAC
Extragalactic Database, NED), for which 1\arcsec\ corresponds to 296\,pc at the
galaxy.  The near-IR gas excitation and kinematics as well as the stellar
kinematics of the inner 450 pc of Mrk\,1157 are discussed in Riffel \&
Storchi-Bergmann ({\it in prep.}) using the same data used here.

This paper is structured as follows. In Sec.~\ref{data} we describe the
observations, data reduction and cleaning procedures. The spectral fitting
procedures are discussed in Sec.~\ref{synt}. In Sec.~\ref{results} we present our
results, which are discussed in Sec.~\ref{discussion}.  The conclusions are
presented in Sec.~\ref{conclusions}.

\section{The data}\label{data}

\subsection{Observations and Data Reduction}

Mrk\,1157 was observed with Gemini NIFS \citep{mcgregor03} operating with the
Gemini North Adaptive Optics system ALTAIR in September/October 2009 under the
programme GN-2009B-Q-27, following the standard Object-Sky-Sky-Object dither
sequence, with off-source sky positions since the target is extended, and
individual exposure times of 550\,s.

Two sets of observations with six on-source individual exposures were obtained at
different spectral ranges: the first at the J-band, centred at 1.25\,$\mu$m and
covering the spectral region from 1.14\,$\mu$m to 1.36\,$\mu$m, and the second
at the K$_{\rm l}$-band, centred at 2.3\,$\mu$m, covering the spectral range
from 2.10$\,\mu$m to 2.53$\,\mu$m.  At the J-band, the J\_G5603 grating and
ZJ\_G0601 filter were used, resulting in a spectral resolution of
$\approx1.8\,\AA$, as obtained from the measurement of the full width at half
maximum (FWHM) of ArXe arc lamp lines.  The K$_{\rm l}$-band observations were
obtained using the Kl\_G5607 grating and HK\_G0603 filter and resulted in a
spectral resolution of FWHM$\approx3.5\,\AA$.

The data reduction was accomplished using tasks contained in the {\sc nifs}
package which is part of {\sc gemini iraf}\footnote{IRAF is distributed by the
  National Optical Astronomy Observatories, which are operated by the
  Association of Universities for Research in Astronomy, Inc., under cooperative
  agreement with the National Science Foundation.} package, as well as generic
{\sc iraf} tasks. The reduction procedure included trimming of the images,
flat-fielding, sky subtraction, wavelength and s-distortion calibrations. We
have also removed the telluric bands and flux calibrated the frames by
interpolating a black body function to the spectrum of the telluric standard
star.

The angular resolution obtained from the FWHM of the spatial profile of the
telluric standard star is 0\farcs11$\pm$0\farcs02 for the J-band and
0\farcs12$\pm$0\farcs02 for the K$_{\rm l}$-band, corresponding to 32.6$\pm$5.9
and 35.5$\pm$5.9\,pc at the galaxy, respectively.  Since the star has been observed for a 
shorter time than the galaxy, this value should be considered as a lower limit for the spatial resolution. Nevertheless, the performance 
of ALTAIR for larger integration times (such those used for the galaxy) in previous works by our group was found to be similar to the one reached in the observation of the star. In the 
case of NGC\,4151, the spatial resolution of the galaxy is only 0\farcs02 worse than those obtained from the FWHM of the brightness profile 
of the star \citep{sb09}. For each band, individual datacubes were created at an angular sampling of 0\farcs05$\times$0\farcs05 and
combined to a single datacube using the {\sc gemcombine iraf} task. More
details about observations and data reduction can be found in Riffel \&
Storchi-Bergmann ({\it in prep.}).

\subsection{Removing noise effects: The WPCA Method}

Because of the complexity of the instruments, hyperspectral data frequently
contains instrumental signatures that cannot be removed with standard data
reduction techniques, and which in many cases contaminates the science data.
The Gemini NIFS is no exception. The wavelet principal component analysis (WPCA)
technique described below (a full account of the technique will be addressed in
a paper in preparation - Ferrari et al. 2011) can separate the
instrumental fingerprint and, after its removal, allow the analysis of the ``clean"
data.

\subsubsection{Principal Component Analysis}
Besides the relevant information, any set of data has some amount of redundant
information and noise. The goal of principal component analysis (PCA) is to find
a new basis on which the data -- the spectra from each spatial pixel - is expressed in a more meaningful way.  The basis
vectors -- \textbf{the principal components} -- are obtained by searching for the maximum variance in the data. 
There are as many vectors as the original variables, but
usually the first few vectors contain most of the information of the data set:
they provide a more concise description of the data. Often the physical content
is restricted to few vectors and thus interpretation can be easier.
Mathematically we proceed as follows. In a basis where the variables are all
uncorrelated (orthogonal) their covariance or correlation matrix is diagonal. We
then find the principal components by diagonalizing the covariance or
correlation matrix of the original data. In the case of covariance matrix, the
\textbf{eigenvectors} are the principal components and the \textbf{eigenvalues}
are the variance associated with each of the eigenvectors. In the case of a
hyperspectral cube, it is informative to measure the correlation between each
principal component and the spectra in each spatial pixel: the
\textbf{tomograms}. For example, tomogram 1 is the projection (scalar product)
of eigenvector 1 and the spectra relative to each spatial pixel.  For a full
description on the method applied to astronomical data, see \citet{steiner09}.

\subsubsection{Discrete Wavelet Transform}
The Discrete Wavelet Transform (DWT) consists of describing a signal $C_0$ in
terms of a smoothed component $C_J$ and detail wavelet coefficients $\{W_j\}$
\citep{mallat99,stark06}:
\begin{equation}
\label{eq:dwt}
C_0 = C_J + \sum_{j=1}^{J} W_j.
\end{equation}
The maximum wavelet scale is $J$.  In the \emph{\`a trous} (i.e. with holes)
algorithm \citep{stark06} used in this work, a series of smoothed versions
$\{C_j\}$ of $C_0$ are calculated by convolving $C_0$ with scaled versions of a
low-pass filter $h$. In this way, the term $C_j$ is calculated by taking
adjacent pixels from $C_{j-1}$ that are $2^j$ pixels far apart. We proceed in this way
for each of the scale $j=1\ldots J$. After this, the discrete wavelet transform
is obtained from the difference
\begin{equation}
  \label{eq:wdiff}
  W_j = C_j-C_{j-1}.
\end{equation}
The wavelet coefficients $\{W_j\}$ now retain only details which are about $2^j$ in
size, a multiresolution transform. The maximum scale $J$ is mathematically
arbitrary because, by virtue of Eq.~(\ref{eq:dwt}) and (\ref{eq:wdiff}),
reconstruction is complete for any $J$. Physically, $J$ is chosen so that the
spatial frequencies of interest are lower than $2^J$.

\subsubsection{The WPCA Algorithm} 
\begin{figure}
\centering \includegraphics[width=.5\textwidth]{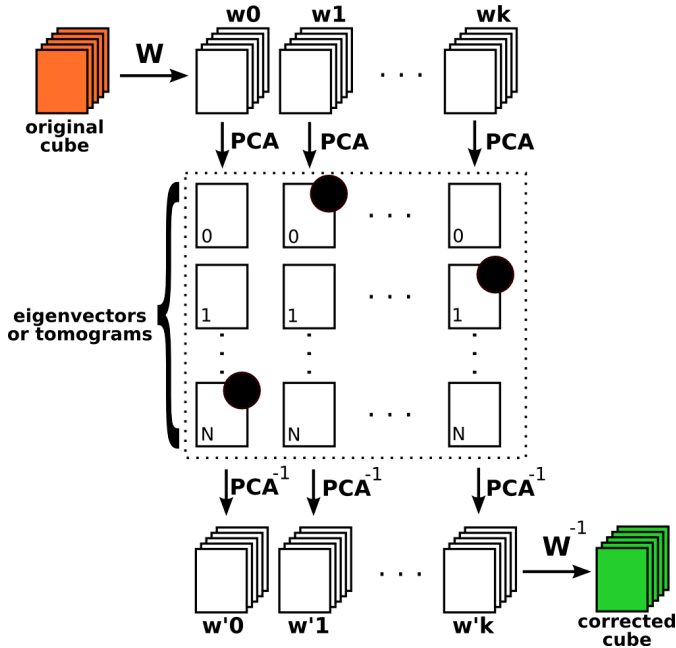} 
\caption{Wavelet and PCA combined: each spatial frame of the original datacube
  is first decomposed in wavelet ($W$) space and then the PCA analysis is
  performed. After identifying and correcting for unwanted effects (instrumental
  signatures, fingerprints or noise, for example) the cube is reconstructed in
  both PCA and wavelet spaces. }
\label{fig:wpca} 
\end{figure}

The idea behind the use of the wavelet and the PCA transforms together is to
decompose the original signal both in the wavelet and in the PCA space, select
unwanted features and finally reconstruct the signal after eliminating these
features. The algorithm is schematically shown in Figure \ref{fig:wpca}. The
original datacube (orange) is first decomposed in wavelets scales $W_0,
W_1,\ldots,W_k$. Then, the PCA analysis is performed in each of the wavelet
components, resulting in several eigenvectors for each scale. After removing
those eigenvectors which contain unwanted structures (represented by black dots
in the figure), each of the wavelets components is restored with the PCA
reconstruction, $PCA^{-1}$, and then the resulting corrected wavelets $W'_0,
W'_1,\ldots,W'_k$ are combined to form the resulting corrected cube (green).

\subsubsection{Application to Mrk\,1157}

\begin{figure*}
\includegraphics[width=\textwidth]{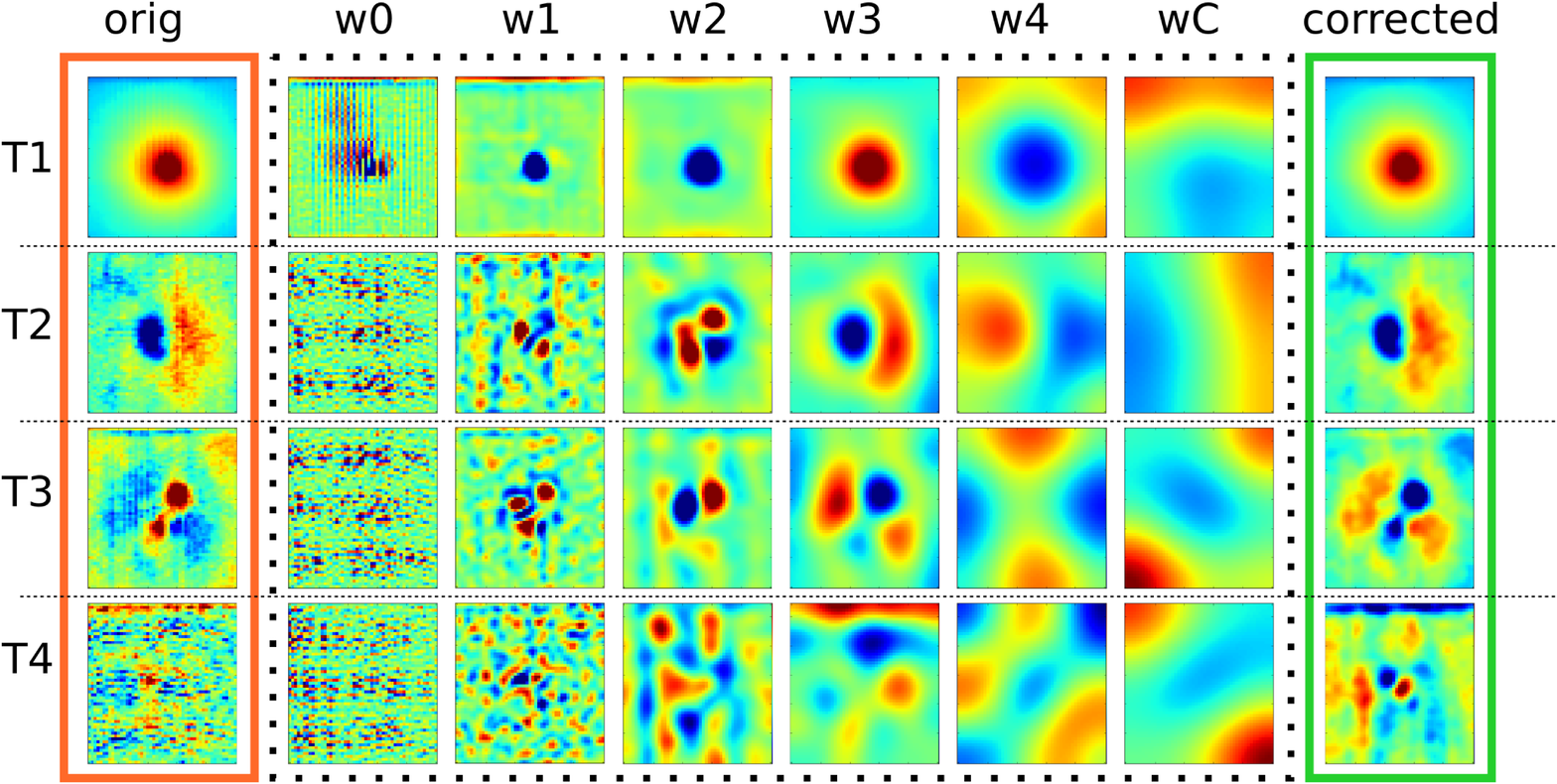} 
\caption{ WPCA decomposition of the Mrk\,1157 $J$- band datacube. Tomograms are separated in rows,
  from top to bottom, tomograms 1 to 4. Original data (orange rectangle), is wavelet
  decomposed (black dotted-rectangle) in 5 scales plus continuum -- $w_i,\ i=0\ldots 4$ and
  $w_C$ -- and then reconstructed without the first component $w_0$ (green
  rectangle). Colors in individual images are false to enhance details. The first 4
  tomograms shown correspond to $\sim 95\%$ of the variance of the data
  cube. Figure frames have the same color (rectangles) scheme as in Fig~\ref{fig:wpca}.  Color may appear inverted because tomograms, as well as
eigenvectors, are determined from the PCA except for its sign.}
\label{pcaNGC591} 
\end{figure*}

Figure ~\ref{pcaNGC591} shows an example of the WPCA transform of Mrk\,1157 datacube. The
columns are the wavelet decomposition of the original cube. In both
Fig.\ref{fig:wpca} and \ref{pcaNGC591} the dotted rectangle represents the same
kind of information: in each wavelet scale a PCA analysis is performed. The original
information is completely represented in the wavelet space, thus summing up all
the components gives back the original data.  The WPCA cleaning process is
basically to identify the components (in wavelet and PCA space) which contains
mostly instrumental signatures (i.e noise), then remove them and reconstruct the
signal.

In the case of Mrk\,1157 observations, the WPCA decomposition reveals that the
$W_0$ tomograms are dominated by the instrumental signature (horizontal and
vertical stripes in Fig.~\ref{pcaNGC591}). That instrumental fingerprint can
also be identified in the eigenvectors as high frequency bumps (not shown) in
the beginning and in the end of the spectra -- the instrumental signature has a
spatial and spectral behaviour. Thus, we remove $W_0$ and reconstruct the cube
with $W_1\ldots W_4, W_C$. We call the attention to the fact that we have
analysed 5 wavelet scales and 20 principal componentes (with 20 eigenvectors and
20 tomograms; a total of 200 maps) to identify the instrumental signature.  For
sake of brevity, in Fig.~\ref{pcaNGC591} we only show the first 4 principal
components, which account for $>$95\% of the data variance.

The {\it clean} datacube was used to perform the stellar population analysis.
The spatial resolution of the J and K$_l$ bands are very similar, allowing us to
combine both datacubes into a single datacube and then perform the spectral
synthesis.  The combination was done using the {\it scombine} {\sc iraf} task at
a spectral sampling of 2\,\AA. The astrometry was done using the peak of the
continuum emission in both bands.  We have re-binned the resulting datacube to
a spatial sampling of 0\farcs1 in order to reach a signal-to-noise ratio (S/N)
high enough to obtain a reliable fit of the stellar population. The resulting
S/N ranges from $S/N\approx20$ at the borders of the NIFS field to up to 120 at
the nucleus.  We used the {\it wspectext} {\sc iraf} task to convert the
individual spectra of the final datacube to 784 {\it ascii} files, used as
input to the synthesis code. The observation field is 2\farcs8$\times$2\farcs8,
which corresponds to 830$\times$830 pc$^2$ at the galaxy.

\section{Spectral Synthesis}\label{synt}

Our main goal in this section is to map in detail the NIR spectral energy
distribution (SED) components of the inner 400 pc of Mrk\,1157.  For this
purpose we fit simultaneously the underlying continuum of the $J$ and $K$ bands
applying the method described in \citet{rogerio09} and summarized below.

In order to perform the stellar population synthesis we use the \str\ code
\citep{cid04,cid05a,mateus06,asari07,cid08}. \str\ fits the whole underlying
spectrum, excluding emission lines and spurious data, mixing computational
techniques originally developed for semi empirical population synthesis with
ingredients from evolutionary synthesis models
\citep{cid04,cid05a}. Essentially, the code fits an observed spectrum,
$O_{\lambda}$, with a combination, in different proportions, of $N_{\star}$
single stellar populations (SSPs).  Extinction is parametrised by the V-band
extinction $A_V$ and modelled by \str\ as due to foreground dust.  In the fits
we use the CCM \citep{cardelli89} extinction law.  In order to model a spectrum
$M_{\lambda}$, the code solves the following equation:

\begin{equation}
M_{\lambda}=M_{\lambda 0}\sum_{j=1}^{N_{\star}}x_j\,b_{j,\lambda}\,r_{\lambda}  \otimes G(v_{\star},\sigma_{\star})
\label{streq}
\end{equation}
where {\bf x} is the population vector, whose components $x_j$,  (j = 1,..,N$\star$) represent the fractional contribution 
of each SSP in the base to the total synthetic flux at $\lambda_0$. $b_{j,\lambda}$ is the spectrum of the
$j$th SSP of the base of elements normalized at $\lambda_0$, the reddening term
is represented by $r_{\lambda}=10^{-0.4(A_{\lambda}-A_{\lambda 0})}$,
$M_{\lambda 0}$ is the synthetic flux at the normalisation wavelength, $\otimes$
denotes the convolution operator and $G(v_{\star},\sigma_{\star})$ is the
Gaussian distribution used to model the line-of-sight stellar motions, which is
centred at velocity $v_{\star}$ with dispersion $\sigma_{\star}$.  However, note that due to the low spectral resolution 
of EPS models in the NIR, the $\sigma_{\star}$ values cannot be derived in a reliable way from the synthesis, thus, we do not use them. For details on $\sigma_{\star}$ see Sec.~\ref{results}.  The final fit
is carried out minimizing the equation:

\begin{equation}
\chi^2 = \sum_{\lambda}[(O_{\lambda}-M_{\lambda})w_{\lambda}]^2
\end{equation}
where emission lines and spurious features are excluded from the fit by fixing $w_{\lambda}$=0. 

In Eq.~\ref{streq}, the most important ingredient in stellar population
synthesis is the base set, $b_{j,\lambda}$. As default base \str\ uses the SSPs
of \citet{bc03}.  However, these SSPs do not include the effect of the
contribution of thermally pulsating asymptotic giant branch (TP-AGB) stars,
whose contribution is enhanced in the NIR and crucial to model the stellar
populations in this spectral region
\citep[see][]{rogerio07,rogerio08,rogerio09,rogemar10b,martins10}.  Thus, we
update the base using the \citet{maraston05} Evolutionary Population Synthesis
(EPS) models as described in \citet{rogerio09}. The base comprises SSPs synthetic
spectra covering 12 ages ($t=$0.01,0.03, 0.05, 0.1, 0.3, 0.5, 0.7, 1, 2, 5, 9
and 13~Gyr) and four metallicities ($Z=$0.02, 0.5, 1, 2~Z$_\odot$).  We also
include black-body functions for temperatures in the range 700-1400\,K in steps
of 100\,K \citep{rogerio09} and a power-law ($F_\nu\propto\nu^{-1.5}$) in order
to account for possible contributions from dust emission ($BB$) and from a
featureless continuum ($FC$), respectively, at the nucleus
\citep[e.g.][]{cid04}. The same spectral base was used to map the age
distribution of the stellar population in the inner 300 pc of Mrk\,1066
\citep{rogemar10b}.

\section{Results}\label{results}

\begin{figure*}
  \centering
  \includegraphics[scale=0.85]{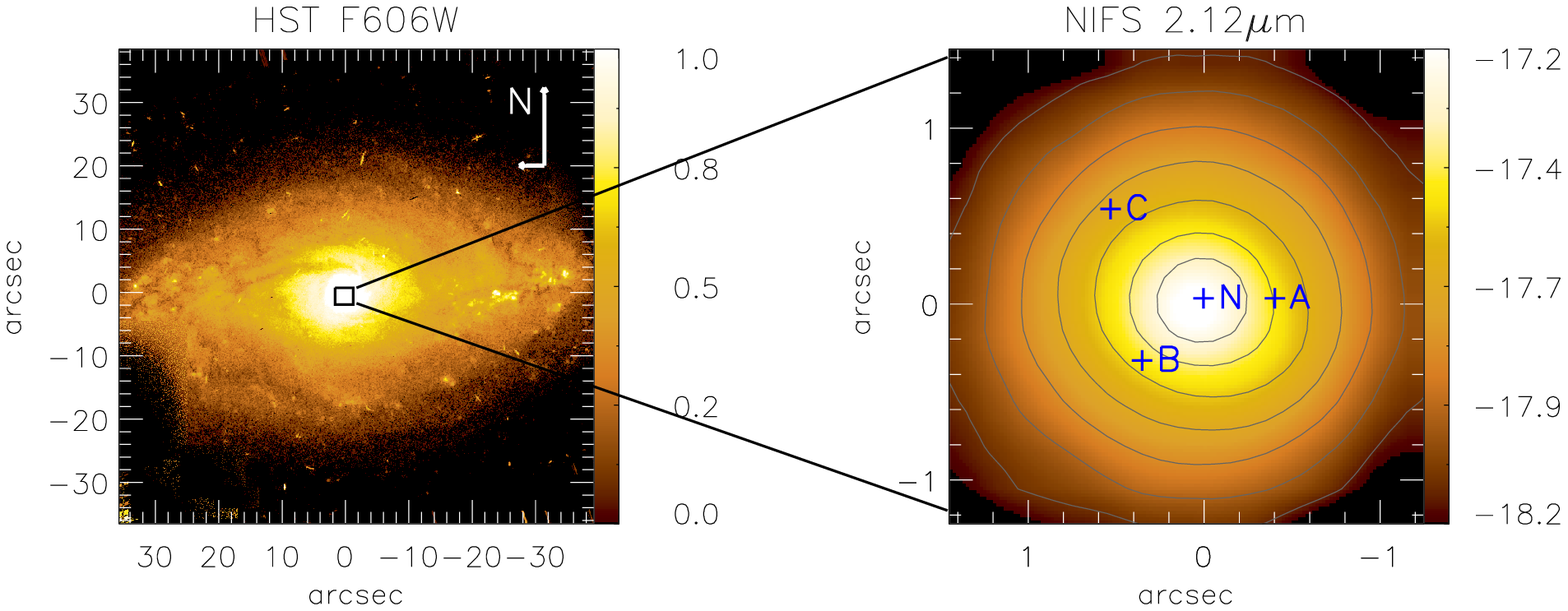} 
  \includegraphics[scale=0.8]{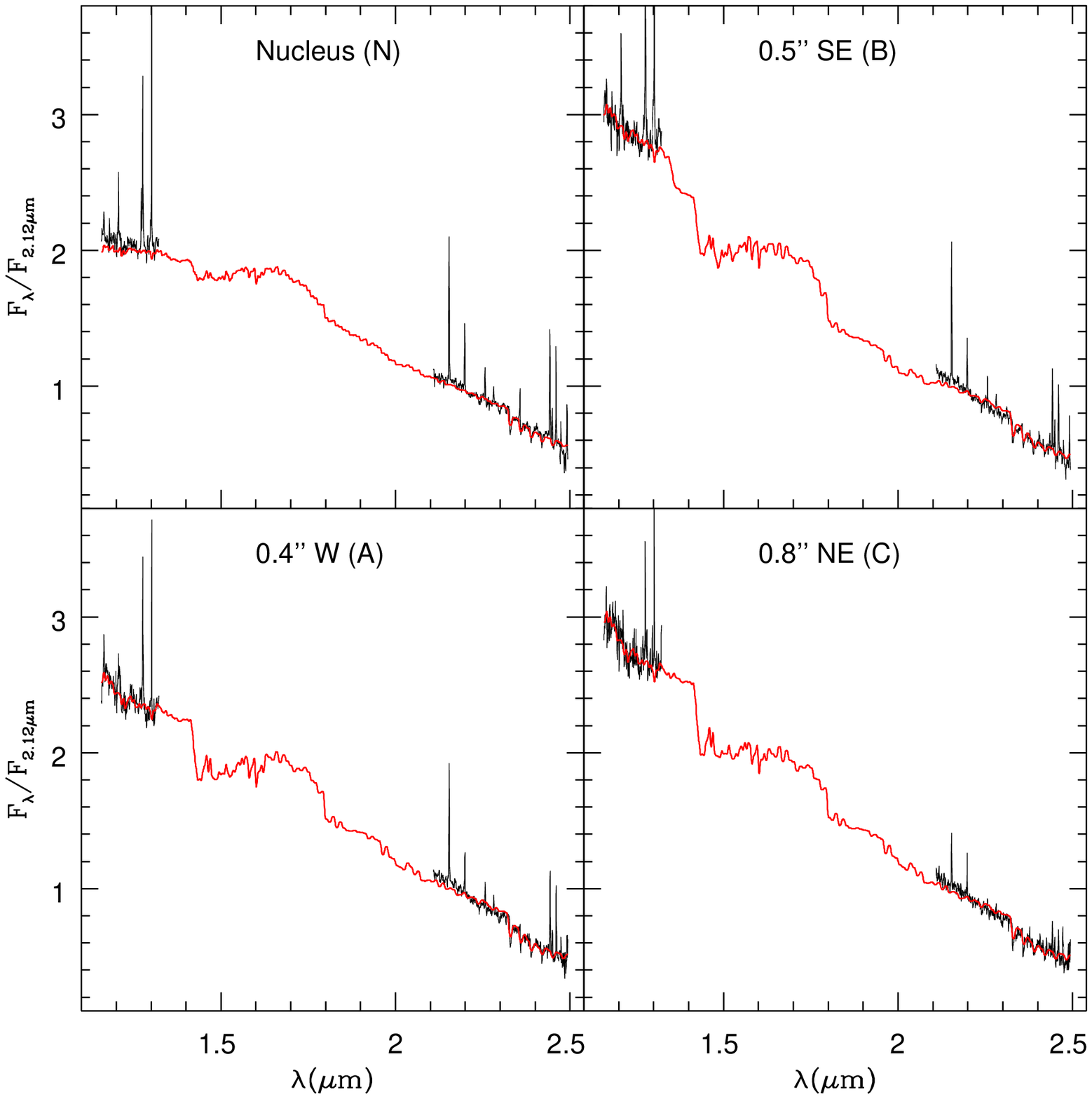} 
  \caption{Top-left panel: HST WFPC2 continuum image of Mrk\,1157 obtained
    through the filter F606W \citep{malkan98}. Top-right panel: 2.17\,$\mu$m
    continuum image obtained from the NIFS datacube. Bottom panels show typical
    spectra obtained within an 0\farcs25$\times$0\farcs25 aperture for the
    nucleus and for a location at 0\farcs4 north-west from it (position A). The
    box in the HST image shows the NIFS field of view. Bottom panels show
    typical spectra obtained within an 0\farcs1$\times$0\farcs1 aperture and the
    resulting fit of the stellar population for the positions marked at the
    top-right panel.}
  \label{large}  
\end{figure*}

In Fig.~\ref{large} (top-left) we show an optical image of Mrk\,1157 taken with
the Wide Field and Planetary Camera 2 (WFPC2) at the Hubble Space Telescope
(HST) through the filter F606W \citep{malkan98}. The 2.17\,$\mu$m NIFS datacube
continuum image is also shown in Fig.~\ref{large} (top-right).  In order to
illustrate the accuracy of our fits, we show in Fig.~\ref{large} (bottom) sample
spectra, obtained within 0\farcs1$\times$0\farcs1 apertures for four distinct
positions: the nucleus (position N marked at top-left panel of Fig.~\ref{large})
and 0\farcs4 west (position A), 0\farcs5 south-east (position B) and 0\farcs8
north-east (position C) of the nucleus. The synthetic spectra were overploted on
the data as dotted lines. As can be observed in this figure, the modelling of
the stellar population reproduces very well the continuum/absorption spectra at
all positions of Mrk\,1157.  Following \citet{rogemar10b} the observed and
synthetic spectra were normalized at 2.12\,$\mu$m, a region free of
emission/absorption lines \citep{rogerio08}.

\begin{figure}
\centering
\includegraphics[scale=0.45]{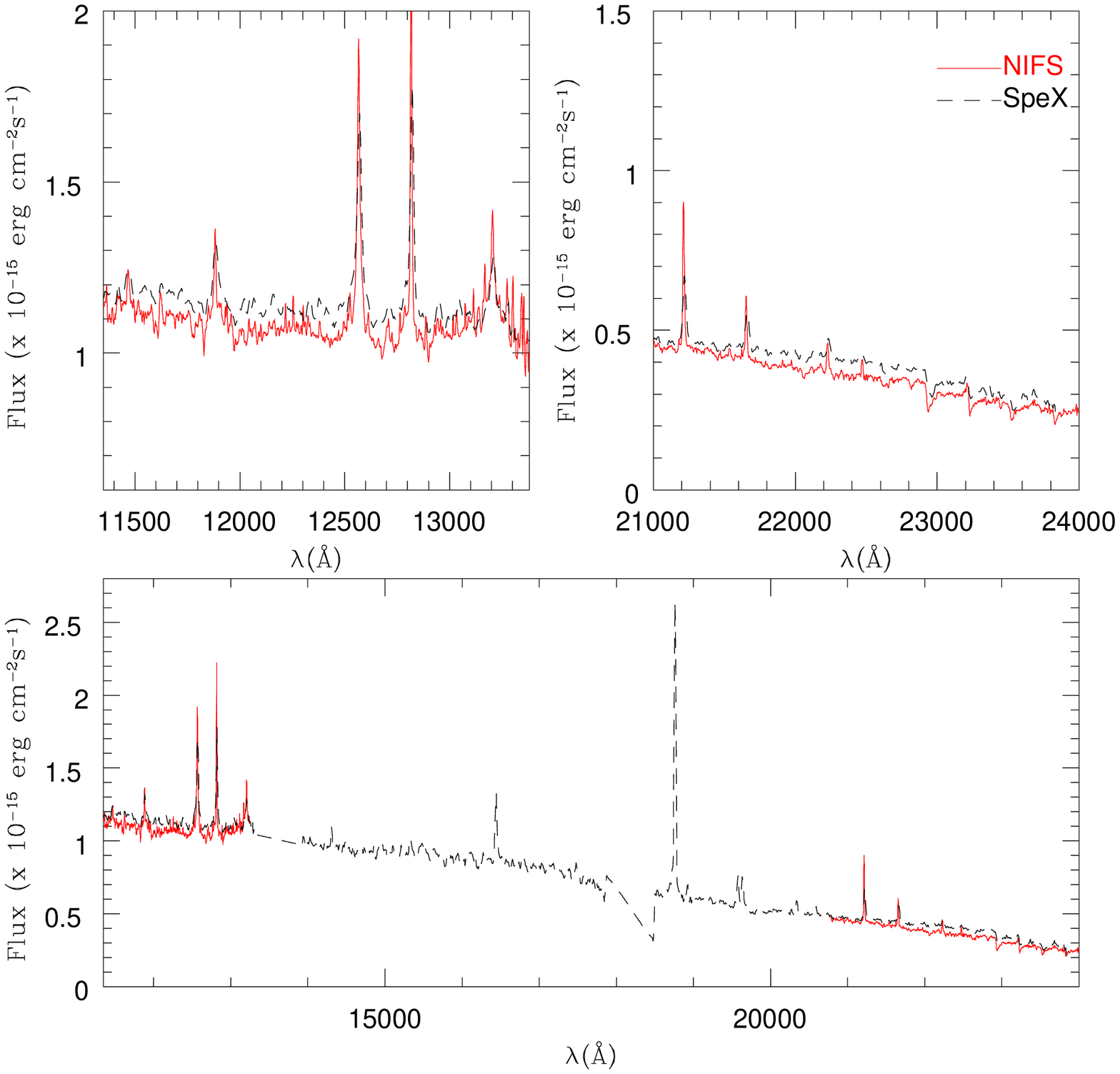} 
\caption{Comparison of the overall shape of our NIFS spectra with the SpeX cross dispersed spectrum of \citet{rogerio06}. }
\label{nifsxspex}  
\end{figure}

No contribution of the $FC$ and/or $BB$ components was necessary in order to
reproduce the the nuclear continuum in Mrk\,1157. Similar result was obtained by
\citet{rogerio09} when fitting the nuclear integrated spectrum of this source within a larger aperture
(r$\simeq$200\,pc) over the 0.8---2.4\mc\ spectral region. We point out that the
synthetic spectra was reddened using the A$\rm_V$ derived from the synthesis
(Fig.~\ref{pop2}).

\begin{figure*}
\centering
\includegraphics[scale=0.65]{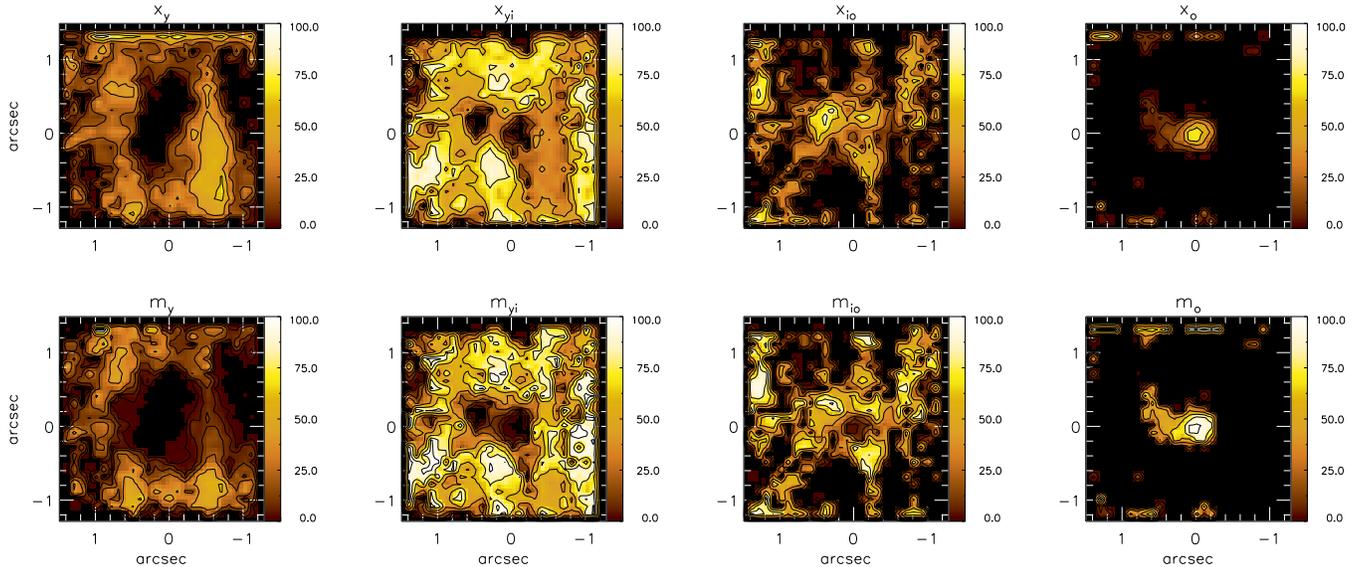} 
\caption{Spatial distributions of the percent contribution of each SPC to the flux at
  $\lambda=2.12\,\mu$m ($x_j$) and to the mass ($m_j$), where $j$ represents the
  age of the SPC: young ($y$: $\leq$ 100 Myr), young-intermediate ($yi$:
  0.3--0.7~Gyr), intermediate-old ($io$: 1--2~Gyr) and old ($o$: 5--13~Gyr).}
\label{pop1}  
\end{figure*}

Following \citet[][see also Riffel et al., 2009 and Cid-Fernandes et
al. 2004]{rogemar10b}, we have binned the contribution of the SPCs ($x_j$) into
a reduced population vector with four age ranges: {\bf young}
($x_y$: $t \leq 100$~Myr); {\bf young-intermediate} ($x_{yi}$: $0.3 \leq t \leq
0.7$~Gyr), {\bf intermediate-old} ($x_{io}$: $1 \leq t \leq 2$~Gyr) and {\bf
  old} ($x_{o}$: $5 \leq t \leq 13$~Gyr).  In Fig.~\ref{pop1}, we show the
spatial distribution of the percent flux contribution at 2.12\,$\mu$m of the
stars in each $\vec{x}$: while in regions farther than r$\sim$ 0.8(240\,pc) from
the nucleus the contribution of the young-intermediate SPCs reaches values of up
to 100\%, closer to the nucleus the contribution of this component is
negligible.  Within this region the stellar population is dominated by the old
component($\approx$50\%), closely surrounded (r $\lesssim$ 0\farcs4) by an intermediate-old
population (also $\approx$50\%). No signs of young
stellar populations were detected close to the nucleus; however, a significant
contribution (25-60\%) is found in regions farther than $r\sim$0\farcs6.

The light-fraction SPC contributions depend of the normalization wavelength and
thus the comparison with results from other spectral regions should be done with
caution \citep{rogerio10}. However, a physical parameter which does not depend
on the normalization point used in the fit is the stellar mass. The mass-fraction of
each population vector components is show in Fig.~\ref{pop1} (young: $m_y$,
young-intermediate: $m_{yi}$, intermediate-old: $m_{io}$ and old population:
$m_o$). The maps of the mass-weighted SPC follow a similar distribution to the
light-weighted ones. However, in the former the contribution of the older ages
is enhanced, particularly within 0\farcs4 from the nucleus.

Besides the SPC distributions, the \str\ outputs the average reddening of the
stellar populations (Fig.\,\ref{pop2}). The highest values of $E(B-V)=0.7$ (we used $\rm A_V$=3.1E(B-V)) are
reached at the nucleus up to within $\sim$0\farcs2 from it.

The goodness of the fit is measured in \str\ by the percent mean deviation:
\textit{adev}=$|O_\lambda-M_\lambda|/O_\lambda$, where $O_\lambda$ is the
observed spectrum and $M_\lambda$ is the fitted model \citep{cid04,cid05}.  The
\textit{adev} map for Mrk\,1157 is shown in Fig.~\ref{pop2} and presents values
$adev\lesssim5$\,\% at most locations, indicating that the model reproduces very
well the observed underlying spectra.

We point out that, in order to have a robust result on stellar population fitting with \str\ it is important to have a 
reliable flux calibration (see \str\ manual). In other words the fit depends on the overall shape of the 
observed spectrum from the $J$ to the $K$-band, and our NIFS observations misses the $H$-band. In order to verify the reliability of our flux calibration we have extracted 
a spectrum from our datacubes matching the aperture and position angle of our SpeX cross-dispersed spectrum \citep{rogerio06}. Note that the SpeX data were taken in the 
cross-dispersed mode, and thus are free from seeing and aperture effects \citep[see][for details]{rogerio06}. The comparison between both spectra is shown in Fig.~\ref{nifsxspex}. It is clear that our NIFS 
spectra have a reliable flux calibration. In fact the difference between the NIFS and SpeX spectra is lower than 3\% in the $J$ band. In addiditon, we have performed 
a simulation varying the flux of the $J$ band by 20\% (from 90\% up to 110\%) and the difference to the population vector components is, in mean, $\lesssim$5\% indicating 
that our results are robust.

\section{Discussion}\label{discussion}

\begin{figure}
\centering
\includegraphics[scale=0.65]{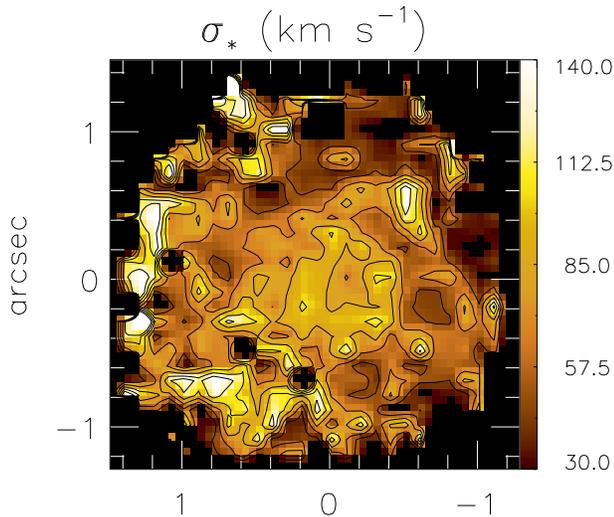} 
\caption{Stellar velocity dispersion map from Riffel \& Storchi-Bergmann (in prep.).}
\label{sig} 
\end{figure}

In general, our results are similar to those obtained in the 2D mapping of the
SP of Mrk\,1066 \citep{rogemar10b} as well as for previous NIR studies using single
aperture nuclear spectra \citep{rogerio09}.  By mapping the stellar population
in 2D we can analyse the spatial variations of the SPCs in the inner few hundred
parsecs of the galaxy. Further, the significance of these variations is enhanced
by the comparison with the $\sigma_*$ map presented in Fig.~\ref{sig}. This map was obtained by fitting the K-band CO absorption
band-heads with the penalized Pixel Fitting (pPXF) method of
\citet{cappellari04} using as stellar template spectra those of the Gemini
library of late spectral type stars observed with the Gemini Near-Infrared
Spectrograph (GNIRS) IFU and NIFS \citep{winge09}. More details on the stellar
kinematics of the central region of Mrk\,1157 can be found in Riffel \&
Storchi-Bergmann({\it in prep.}). The $\sigma_*$ map shows a partial
ring of low-$\sigma_*$ values ($\approx$ 50-60\,${\rm km\,s^{-1}}$) surrounding
the nucleus at $\approx$ 0\farcs8 (230\,pc) from it, immersed in higher
$\sigma_*$ values of the bulge stars ($\approx$ 100\,${\rm km\,s^{-1}}$).

Such rings are commonly observed in the central region of active galaxies and
are due to kinematically colder regions with younger stars than the underlying
bulge \citep{barbosa06,deo06,lopes07,rogemar08,rogemar09a}.  The comparison
between Mrk\,1157 stellar population synthesis maps (light- and mass-weighted)
and the $\sigma_*$ map shows that the low $\sigma_*$ ring is spatially
correlated with the young-intermediate age SPC, while the highest $\sigma_*$s
are associated with the old component.  Interestingly, very similar results were
found by \citet{rogemar10b} for Mrk\,1066. Thus, the results found for Mrk\,1157
and Mrk\,1066 support the use of low stellar velocity dispersion as a tracer of
young-intermediate age stars in the galaxy centre, thus confirming the
interpretation of the above studies.

The most compressed form to represent the SPCs mixture of a galaxy is by its
mean stellar age, which was defined by \citet{cid05} in two ways: the first is
weighted by light fraction,
\begin{equation}
\langle {\rm log} t_{\star} \rangle_{L} = \displaystyle \sum^{N_{\star}}_{j=1}
x_j {\rm log}t_j, 
\end{equation}
and the second, weighted by the stellar mass, 
\begin{equation}
\langle {\rm log} t_{\star} \rangle_{M} = \displaystyle \sum^{N_{\star}}_{j=1}
m_j {\rm log}t_j. 
\end{equation}
Being both definitions limited by the age range used in our base, the former
is more representative of the younger ages while the latter is enhanced by the
old SPC \citep{rogerio09}.  In order to compare our results with those derived
using integrated spectra we have calculated the mean ages for the central region
of Mrk\,1157. These mean ages in light- and mass-fractions are $\langle {\rm
  log} t_{\star} \rangle_{L}$ = 8.48 $\langle {\rm log} t_{\star} \rangle_{M}$ =
8.65, respectively.  Note that these values were obtained taking the mean value
over all the spectra in the map (e.g. summing $\langle {\rm log} t_{\star}
\rangle_{L}$ values over the map and dividing the result by the number of
spectra in the map).

Besides our study of Mrk\,1066 \citep{rogemar10b}, to date, the only previous 2D
stellar population studies of active galaxies in the NIR are those from the
group of R. I. Davies. \citet{davies07} investigated the circumnuclear star
formation in nine Seyfert galaxies using the NIR adaptive optics integral field
spectrograph {\sc sinfoni} at the Very Large Telescope (VLT). Based on measurements of the \br\ emission-line
equivalent width, supernova rate and mass-to-light ratio, these authors found
circumnuclear disks of typical diameters of tens of pc with a ``characteristic"
age in the range 10--300~Myr.  Such a ``characteristic" age can be associated
with our mean age ($\sim$ 300 Myr), thus  our result for Mrk\,1157 is in good
agreement with those found by \citet{davies07} for their sample (which does not
include Mrk\,1157).  Nevertheless, as pointed out by \citet{rogemar10b} the
methodology adopted here allowed us not only to obtain a ``characteristic" age,
but to also map the spatial distribution of the SPCs of different ages in the
central region of a Seyfert galaxy, on the basis of NIR integral-field
spectroscopy. Moreover, our results are in reasonable agreement with those of
\citet{rogerio09} where the authors use integrated long-slit spectroscopy
(considering the different apertures).

\begin{figure*}
\centering
\includegraphics[scale=1]{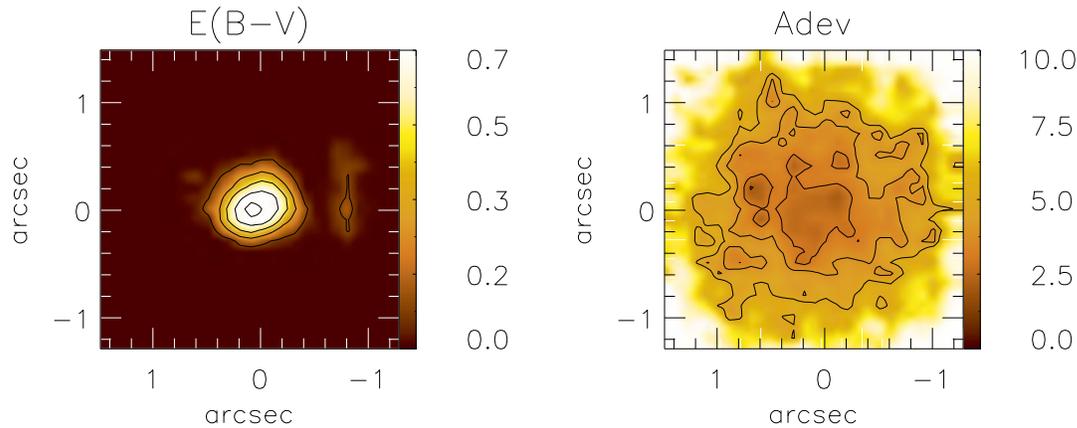} 
\caption{Reddening map and {\it adev} map (percent mean deviation from the spectral fit).} 
\label{pop2}  
\end{figure*}

In further support to the results of the synthesis, we found that the average
reddening map derived for the stellar population (Fig.\,\ref{pop2}) is in close
agreement with the one derived for the narrow-line region using emission-line
ratios (Riffel \& Storchi-Bergmann, {\it in prep.}), presenting a similar
structure and a very similar average value to that found by
\citet[][E(B-V)$\approx$0.63]{rogerio09} for the SP, using NIR integrated
spectra.

\section{Conclusions}\label{conclusions}

As part of an ongoing project aimed at mapping the age distribution of the NIR
SP in the inner few hundred parsecs of Seyfert galaxies, we present the spectral
synthesis for the nuclear region of the Seyfert 2 galaxy Mrk\,1157 (NGC\,591)
within the inner $\sim$ 400 pc at spatial resolution of $\approx$\,35\,pc. We
have used a cleaning method which allowed us to remove the noise effects and
redundant data from the 1D extractions, thus allowing a more accurate analysis
of the SPCs. Using the noise free cube we have mapped the distribution of
stellar population components of different ages and of their average
reddening. The main conclusions of this work are:

\begin{itemize}

\item 

  The cleaning technique presented here has allowed us to perform the spectral
  synthesis in noisy spectra (e.g. borders of the datacube) in a more accurate
  way.

\item The age of the dominant stellar population presents spatial variations:
  the flux and mass contributions within the inner $\approx$\,160~pc are
  dominated by old stars ($t \ge$5\,Gyr), while intermediate age stars ($0.3
  \leq t \leq 0.7$~Gyr) dominate in the circumnuclear region between 170 pc and
  350pc.

\item As for Mrk\,1066 \citep{rogemar10b}, we found that there is a spatial correlation between the
  distribution of the intermediate age component and low stellar velocity
  dispersion values, which delineate a partial ring around the nucleus of
  Mrk\,1157. Similar structures have been found around other active nuclei, and
  our results for Mrk\,1157 (and Mrk\,1066) reveals that these nuclear rings are
  formed by intermediate age stars.

\item No signatures of non-thermal and hot dust components are found in central
  region of Mrk\,1157.

\end{itemize}

\section*{Acknowledgments}
We thank an anonymous referee for helpful suggestions. The {\sc starlight} project 
is supported by the Brazilian agencies CNPq, CAPES and FAPESP and by the France-Brazil CAPES/Cofecub program.
This work is based on observations obtained at the Gemini Observatory, which is
operated by the Association of Universities for Research in Astronomy, Inc.,
under a cooperative agreement with the NSF on behalf of the Gemini partnership:
the National Science Foundation (United States), the Science and Technology
Facilities Council (United Kingdom), the National Research Council (Canada),
CONICYT (Chile), the Australian Research Council (Australia), Minist\'erio da
Ci\^encia e Tecnologia (Brazil) and south-eastCYT (Argentina).  This research
has made use of the NASA/IPAC Extragalactic Database (NED) which is operated by
the Jet Propulsion Laboratory, California Institute of Technology, under
contract with the National Aeronautics and Space Administration.  This work has
been partially supported by the Brazilian institution CNPq.

\label{lastpage}

\end{document}